# Study of uranium oxide milling in order to obtain nanostructured UC$_x$ target

Julien Guillot[a*], Sandrine Tusseau-Nenez[a*], Brigitte Roussière[a], Nicole Barré-Boscher[a], François Brisset[b], Maher Cheikh Mhamed[a], Christophe Lau[a], Sophie Nowak[c,d]

[a] Institut de Physique Nucléaire CNRS/IN2P3 UMR 8608 – Université Paris Sud, F-91406 ORSAY Cedex
[b] ICMMO UMR 8182 CNRS – Université Paris Sud, F-91405 Orsay Cedex
[c] ITODYS, Université Paris Diderot, Sorbonne Paris Cité, CNRS UMR-7086, F-75013 Paris
[d] IPSL CNRS UMR 7583 Universités Paris Est Créteil et Paris Diderot, F-94010 Créteil Cedex
Corresponding author: guillotjulien@ipno.in2p3.fr

___________________________________________________________________________________

___________________________________________________________________________________

**Abstract:**

A R&D program is developed at the ALTO facility to provide new beams of exotic neutron-rich nuclei, as intense as possible. In the framework of European projects, it has been shown that the use of refractory targets with nanometric structure allows us to obtain beams of nuclei unreachable until now. The first parameter to be controlled in the processing to obtain targets with a homogeneous nanostructure is the grinding of uranium dioxide, down to 100 nm grain size. In this study, dry and wet grinding routes are studied and the powders are analyzed in terms of phase stabilization, specific surface area and grain morphology. It appears that the grinding, as well dry as wet, leads to the decrease of the particle size. The oxidation of $UO_2$ is observed whatever the grinding. However, the dry grinding is the most efficient and leads to the oxidation of $UO_2$ into $U_4O_9$ and $U_3O_7$ whose quantities increase with the grinding time while crystallite sizes decrease.

___________________________________________________________________________________

**Introduction**

The ISOL technique is commonly used worldwide for the production of radioactive nuclear beams (RIB). At the ALTO facility, the neutron-rich nuclei are produced by photofission in thick uranium carbide targets that are the subject of a R&D program [1]. To improve the intensities of radioactive beams, in particular those formed by short-lived isotopes, dense and porous uranium carbide target (noted UC$_x$, as usually $UC_2$ is obtained with minor amount of UC) has to be developed. These properties (dense and porous) are a priori antagonistic but necessary to increase respectively the amount of fission fragments produced and their diffusion out of the target. The improvement of such targets requires the control of the microstructure of the material prepared by carburization between uranium dioxide and graphite according to the reaction:

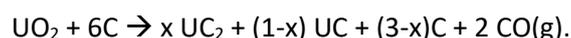

$UO_2 + 6C \rightarrow x\ UC_2 + (1-x)\ UC + (3-x)C + 2\ CO(g).$

Recently, within the framework of SPIRAL2 [2] and the ActILab (ENSAR Transnational Access Activity within the European VII Framework Program) projects, studies were performed in order to correlate the microstructure of UC$_x$ pellets with their fission product releases. These studies lead us to conclude that the release would be optimal if we succeed in controlling the pellet microstructure, in particular by obtaining a homogeneous nanostructure. In order to reduce the grain growth during sintering and get nanoscaled carbides, powders must have nanosized grains with a particle size distribution as narrow as possible. For that, in this work, we study the impact of grinding of $UO_2$ powder, the first key parameter to control.

An alternate way of producing $UC_x$ can be to replace $UO_2$ by $U_3O_8$ powder as its transformation into $UO_2$ during the carburization could produce higher level of open porosity, a major key parameter for the fission products (FPs) release. Some previous tests were performed using $U_3O_8$ instead of $UO_2$ [2]. Actually, the $UC_x$ open porosity was the highest obtained, about 60%, but recently the use of ground $UO_2$ leads to the same level of open porosity and clearly the FPs is correlated to the homogeneity of the $UC_x$ target microstructure (Tusseau-Nenez *et al.*, accepted in NIMB).

**Material and methods**

The uranium dioxide powder (AREVA, sample n°65496) used contains 0.25 wt% of $^{235}U$, with impurities of Cr, Ni and Fe, respectively 4, 6 and 16 µg/gU.
It is generally accepted that an energetic grinding allows reducing the grinding time and increasing the particle fracture. Moreover, in order to perform an optimal grinding and obtain a nanopowder, we need to use different sizes of beads. Indeed, the smaller the bead diameter, the smaller the grain size, for a long milling time and an energetic grinding with a high rotation speed [3].
$UO_2$ was milled with a planetary ball mill (PM100, Retsch) with a 50 mL stainless steel bowl and different diameter stainless beads. Wet grinding was performed using isopropanol (IPA). The rotation speed was fixed at 600 rpm (92% of the maximal speed) in order to obtain an energetic grinding. Batches of 10 g of $UO_2$ were prepared using 16 g of stainless steel beads (3 beads with 9 mm diameter, 9 with 5 mm and 18 with 3 mm). A volume of 0, 3.5 or 10 mL of IPA was added, leading respectively to a dry milling (referred as D), wet milling with a pasty texture (W) and slurry (S). Different grinding times were tested, from 30 min to 480 min.
All obtained powders were systematically characterized. The particle size during milling was controlled using specific surface area (SSA) by nitrogen adsorption at 77 K with the BET technique - Brunauer, Emmett and Teller (ASAP2020, Micromeritics). X-ray diffraction (XRD) patterns were collected by a X-Ray powder Diffractometer (XRD, D8 Advance, Bruker AXS) in a Bragg-Brentano geometry ($\theta-\theta$) equipped with Cu radiation ($K_{\alpha 1}$ = 1.54178 Å). Due to the use of a punctual scintillator detector, a long data acquisition time (~7 h) was needed: from 10 to 90° $2\theta$, 0.02° step size, 6 s/step. Phase identification was performed with the Highscore software (PANalytical) and powder diffraction files (ICDD PDF4+ 2013). The quantitative phase analysis was performed by using the MAUD software [4] in order to take into account the $K_\beta$ line (Ni filter, no monochromator) and using the Crystallography Open Database [5]. The instrumental resolution was obtained using LaB6 NIST standard (Standard Reference Material 660a, cell parameter = 0.41569162 nm ± 0.00000097 nm at 22.5 °C). The grain morphology was assessed through scanning electron microscopy- SEM (SIGMA, Zeiss).

**Results and discussion**

*Specific surface area measurement*

Figure 1 presents the SSA measurements for the three kinds of milling. Higher SSA are obtained by dry grinding. Indeed, for both conditions tested with IPA, a plateau is reached after 120 min, while particle size decreases, as the SSA increases, in the dry grinding stage. Even after 8 hours of grinding no plateau is reached, a linear evolution of SSA is observed from 180 min to 480 min. Assuming spherical particles, the grain size is estimated at 82 and 75 nm, after 240 and 480 nm of milling. In our case, the wet milling is most efficient in a pasty mixture than in slurry; this is probably due to a higher probability of impacts between beads and grains in pasty mode. In the case of alumina hydrate, a systematic study of the grinding parameter led to conclude in opposite that the higher the density of the slurry, the higher the specific energy needed to obtain a fixed median size [3].

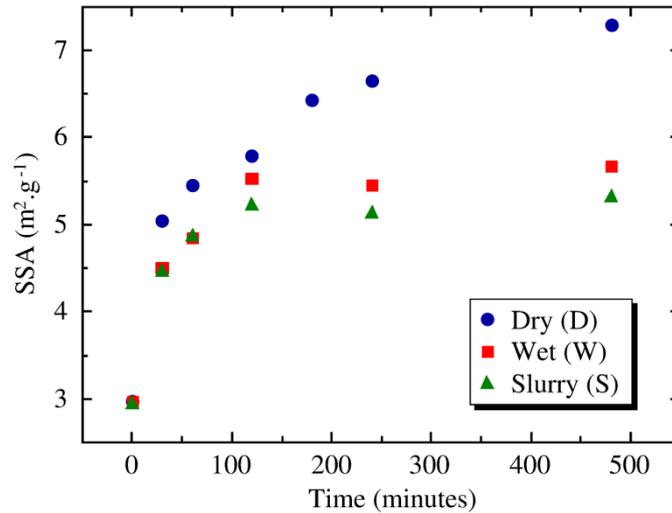

*Figure 1: Specific surface area evolution, for the 3 types of samples, as a function of the milling time. (Error bars ± 0,10 m².g⁻¹)*

*XRD analysis*

Figure 2 presents the XRD patterns for the UO$_2$ raw powder and for the powders ground during 60, 120, 180 and 240 min in dried conditions, in the 25-35° (insert) and 45-60° 2θ ranges.

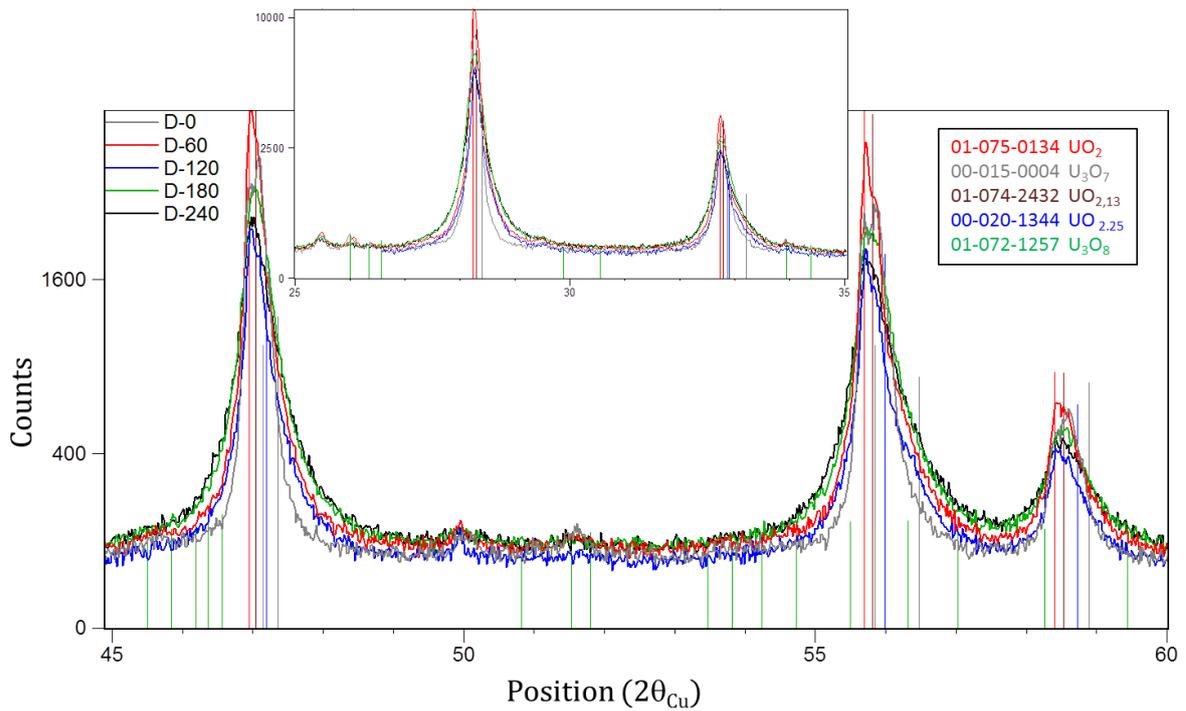

*Figure 2: XRD phase identification for raw and powders ground at different times. Vertical lines correspond to ICDD patterns.*

The grinding leads to the broadening of all the Bragg lines. U$_3$O$_8$ is observed in small quantity in the raw powder. After 180 minutes and more, the peaks are not observed. One can conclude that the grinding leads to a line broadening, then small peaks are drowned in the background. The raw powder exhibits clearly in the 45-60° 2θ range two cubic phases (Fm-3m space group) corresponding to hyperstoichiometric UO$_2$-type phases, UO$_{2+x}$ with 0.10<x<0.25. Whatever the grinding time, the two phases are still observed, even if the line broadening increases with the grinding time. (111) and (222) Bragg peaks are symmetric whereas (200) and (022) are clearly asymmetric indicating a tetragonal structure instead of a cubic one. That indicates the formation of $U_3O_7$ instead of $U_4O_9$ whose intensity of Bragg lines increases in grinding time.

A Rietveld refinement was performed to calculate quantity, crystallite size and microstrains for each phase. The two $UO_2$ type phases (Fm-3m space group) were described by the widely accepted Willis model [6]. Derived from the $UO_2$ structure, the tetragonal $U_3O_7$ phase was described by the cubic $\beta$-$U_4O_9$ phase (F-43m space group) [7], as the $U_3O_7$ crystal contains 16 cuboctahedra per unit cell instead of 12 for the $\beta$-$U_4O_9$ one [8]. In consequence the FWHM of Bragg lines corresponding to tetragonal doublets will be overestimated and will involve crystallite sizes smaller than expected. The crystallite size is calculated to obtain a trend rather than an exact value. Table 1 summarizes the results of the Rietveld refinements. The hyperstoichiometry for $UO_{2+x}$ or $\beta$-$U_4O_9$ was calculated from the equation $a_0 = 5.442 + 0.029\,y$ (for $U_4O_{9-y}$, $0 \leq y \leq 0.31$) [9], according to the phase identification.

| Model | Phase | Crystallographic data | D-0 | D-60 | D-120 | D-180 | D-240 | W-240 | S-240 |
|---|---|---|---|---|---|---|---|---|---|
| $UO_{2+x}$ | $UO_2$ | cell parameter (Å) | 5.4576 ±0.0001 | 5.4596 ±0.0001 | 5.4606 ±0.0001 | 5.4528 ±0.0002 | 5.4579 ±0.0002 | 5.4597 ±0.0004 | 5.4608 ±0.0009 |
| | | quantity (wt%) | 50.8 ±0.7 | 43.4 ±0.4 | 33.8 ±0.4 | 34.4 ±1.0 | 23.6 ±0.7 | 16.7 ±0.8 | 44.3 ±0.5 |
| | | cristallite size (nm) | 170 ±4 | 200♦ ±0 | 200♦ ±0 | 124 ±8 | 221 ±25 | 200♦ ±0 | 214 ±7 |
| | | microstrain | 0.00074 ±0.00002 | 0.00073 ±0.00002 | 0.00084 ±0.00002 | 0.00186 ±0.00003 | 0.00183 ±0.00005 | 0.00048 ±0.00001 | 0.00068 ±0.00001 |
| | | x | 0.14 | 0.12 | 0.11 | 0.19 | 0.13 | 0.11 | 0.10 |
| $UO_{2+x}$ | $U_4O_9$ | cell parameter (Å) | 5.4389 ±0.0002 | 5.4483 ±0.0004 | 5.4493 ±0.0003 | 5.4424 ±0.0004 | 5.4467 ±0.0004 | 5.4463 ±0.0003 | 5.4484 ±0.0003 |
| | | quantity (wt%) | 39.3 ±0.8 | 33.5 ±0.6 | 38.8 ±1.2 | 38.4 ±1.1 | 45.6 ±1.1 | 53.5 ±1.0 | 33.2 ±0.5 |
| | | cristallite size (nm) | 249 ±8 | 409 ±65 | 172 ±32 | 156 ±13 | 106 ±11 | 200♦ ±0 | 200♦ ±0 |
| | | microstrain | 0.00100 ±0.00001 | 0.00450 ±0.00005 | 0.00469 ±0.00007 | 0.00533 ±0.00008 | 0.00580 ±0.00008 | 0.00224 ±0.00004 | 0.00422 ±0.00004 |
| | | x | 0.34* | 0.24 | 0.23 | 0.30 | 0.25 | 0.26 | 0.24 |
| $\beta$-$U_4O_9$ | $U_3O_7$ | cell parameter (Å) | 5.4170 ±0.0000 | 5.4322 ±0.0010 | 5.4399 ±0.0010 | 5.4361 ±0.0008 | 5.4383 ±0.0011 | 5.4395 ±0.0005 | 5.4357 ±0.0008 |
| | | quantity (wt%) | 9.8 ±0.4 | 23.1 ±0.6 | 27.3 ±1.3 | 27.1 ±0.9 | 30.8 ±1.0 | 29.7 ±0.6 | 22.5 ±0.4 |
| | | cristallite size (nm) | 39 ±2 | 11.5 ±0.1 | 14.6 ±0.8 | 13.9 ±0.5 | 11.0 ±0.5 | 25.2 ±0.9 | 10.9 ±0.3 |
| | | microstrain | 0.00008 ±0.0006 | 0.0035 ±0.0006 | 0.0074 ±0.0005 | 0.0088 ±0.0004 | 0.0102 ±0.0004 | 0.00512 ±0.00009 | 0.0034 ±0.0005 |
| | | x | 0.57 | 0.41 | 0.33 | 0.37 | 0.34 | 0.33 | 0.37 |
| | | Rexp | 7.36 | 6.64 | 7.91 | 6.53 | 6.71 | 5.83 | 2.21 |
| | | Rw | 11.13 | 9.33 | 9.55 | 8.04 | 8.06 | 11.47 | 7.1 |
| | | sigma | 1.51 | 1.4 | 1.21 | 1.23 | 1.2 | 1.96 | 3.2 |

*Table 1: Rietveld refinements using Maud software and Delft isotropic model.*
*\*: hyperstoichiometric $U_3O_7$ phase. ♦: fixed crystallite size for a convergent refinement.*

The lattice parameter of $UO_2$ was measured to be 5.470(2) Å [6]. However, the industrial powder is commonly described with a small hyperstoichiometry, about 2.1. Our AREVA powder exhibits three phases: 52 wt% of $UO_{2.14}$, 39 wt% of $UO_{2.34}$ and 9 wt% of $UO_{2.57}$. The correct identification is thus $UO_{2+x}$, $U_3O_7$ and a hyperstoichiometric $U_3O_7$-type phase. Whatever the grinding method, stoichiometric intermediate oxides are stabilized, like $U_4O_9$ and $U_3O_7$, whereas weak variations are observed in the $UO_{2+x}$ stoichiometry.

During the dry grinding, the oxidation occurs with the stabilization, after 180 min of grinding, of $U_4O_9$ and $U_3O_7$ at the expense of the $UO_{2.10}$ phase whose quantity is reduced by a factor 2. The result is in agreement with the description of the oxidation of $UO_2$ reported to occur with a two-steps reaction, leading to the formation of two intermediate oxides $U_3O_7$ and $U_4O_9$, depending on the conditions [10, 11]. The crystallite size of $UO_2$ is not affected by grinding, as we can expected in regards of the Young modulus of $UO_2$ reported as 200-220 GPa [12] with a conchoïdal cleavage, even if a {111} cleavage is not unexpected [13]. Nevertheless, one can suppose that the broadening in this study is due to the reduction of the crystallite size and the increase of microstrain by grinding. Moreover, it was reported that $U_3O_7$ appears as very broad peaks, as observed in this study, explained by a continuous variation of $U_3O_7$ c/a ratio from 1 to 1.03 and also by microstrains due to the mismatch between $UO_2$ and $U_3O_7$ unit cells [11], corroborated by the high level of microstrain calculated herein.

On XRD patterns, the $U_3O_8$ phase is observed with Bragg lines as well intense for W-240 and S-240 as for D-0. This phase is not modified by IPA grinding. After 240 min of pasty milling, $UO_{2.11}$, $U_4O_9$ and $U_3O_7$ are identified. As with the dry grinding, the pasty mode leads mostly to the stabilization of $U_4O_9$ and $U_3O_7$. This method is less efficient in term of crystallite size decrease and in consequence lower level of microstrain is calculated. After 240 min, the slurry milling tends to a rich-$UO_2$ powder and to the thermodynamic stabilization of the $U_3O_7$ metastable phase rather than $U_4O_9$. By this mode, the crystallite size of $U_4O_9$ is slightly decreased, which is consistent with a decrease of the microstrain by this method, as the liquid plays the role of shock absorber.

*Microstructure morphology by SEM*

As shown in Figure 3, on SEM micrographs for 0 and 240 min, the grinding leads to deagglomerate and homogenize the particle size.

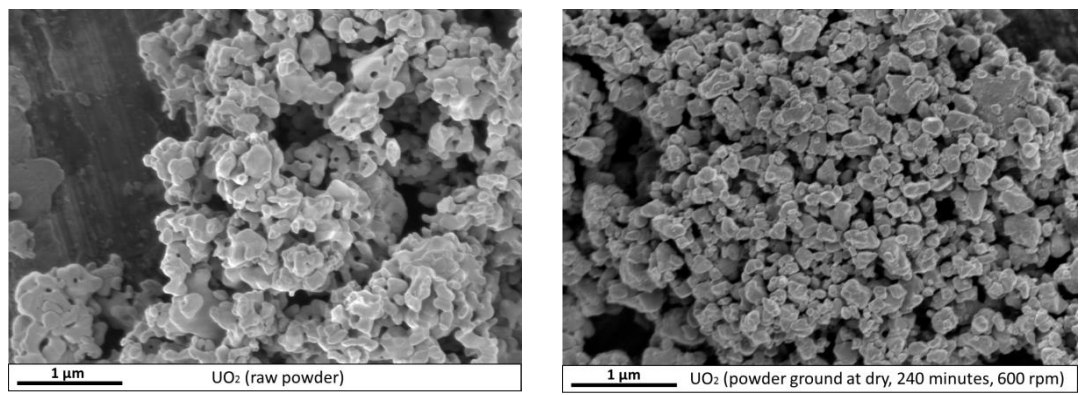

*Figure 3: SEM observation of raw and 240 minutes milled powder*

The average grain size is 200-500 nm before milling and 80-150 nm after 240 minutes. After milling, the angular shape indicates a grain fracture, probably by cleavage. These particle sizes are consistent with the calculated crystallite sizes of $UO_2$ and average values from BET.

**Conclusions**

This study suggests that the UO2 ground particles with an average size, estimated by SEM, from 80 to 150 nm consist of single UO2 crystallites embedded on surface by nanometric hyperstoichiometric uranium oxides, like U4O9 and U3O7 whose particle size decreases with the increase of the milling time.

Clearly, grain size reduction is observed by grinding and the dry route appears more efficient than the IPA one. After 240 minutes of dry milling, and more, $U_4O_9$ and $U_3O_7$ phases are stabilized at the expense of the $UO_2$ phase. Nevertheless, the SSA continues to increase up to 480 minutes of milling, indicating that grain surfaces still expand. This behavior can be correlated to the well-known crack formation in $U_3O_7$ phase, which impacts the mechanical strength of $UO_2$ nuclear fuel. In that case, it was shown that the oxidation leads to the formation of a $U_3O_7$ layer on the $UO_2$ surface grains. In consequence, by milling, the $U_3O_7$ layer is exposed to the impact of beads and cracks can occur easily, as shown by the SSA increase. The pasty grinding leads to the stabilization of the same phases than with the dry one, with similar quantities. The different SSA can be explained by the absorption of the shocks by using IPA. Adding more IPA limits the efficiency of the grinding and reduces the oxidation of the powder, stabilizing mainly $UO_2$. Fixed models were used in our XRD comparative study. To take into account the distortions of the cell due to oxygen insertion into $UO_2$, the latest published model could be used for $U_3O_7$ phase.

In terms of target production, the nanometric size will increase the reactivity of the grain surface of $UO_2$ powder during the carburization. $U_4O_9$ and $U_3O_7$ decomposition into $UO_2$ will promote the degassing responsible for the formation of the open porosity, which is favorable for release properties. The development of new nano-$UC_x$ targets from different nanosized carbon sources (carbon nanotubes, carbon black or graphene) mixed with this nano-$UO_2$ powder will be performed in order to increase the efficiency of radioactive fission product release.